\documentclass[aps]{revtex4}
\usepackage{epsfig}
\newcommand{\ben}{\begin{eqnarray}}
\newcommand{\een}{\end{eqnarray}}
\newcommand{\nnu}{\nonumber\\}
\newcommand{\bef}{\begin{figure}[htb]\centering}
\newcommand{\eef}{\end{figure}}
\newcommand{\sla}[1]{{#1}\!\!\!\slash}

\begin{document}
\title{QCD evolution of naive-time-reversal-odd fragmentation functions}

\author{Zhong-Bo Kang}
\email{zkang@bnl.gov}
\affiliation{RIKEN BNL Research Center,
             Brookhaven National Laboratory,
             Upton, NY 11973, USA}
             
\begin{abstract}
We study QCD evolution equations of the first transverse-momentum-moment of the naive-time-reversal-odd fragmentation functions - the Collins function and the polarizing fragmentation function. We find for the Collins function case that the evolution kernel has a diagonal piece same as that for the transversity fragmentation function, while for the polarizing fragmentation function case this piece is the same as that for the unpolarized fragmentation function. Our results might have important implications in the current global analysis of spin asymmetries.
\end{abstract}

\date{\today}
\maketitle

\section{Introduction}
The phenomenon of single transverse-spin asymmetry (SSA), $A_N\equiv (\sigma(S_\perp)-\sigma(-S_\perp))/(\sigma(S_\perp)+\sigma(-S_\perp))$, defined as the ratio of the difference and the sum of the cross sections when the transverse spin vector $S_\perp$ is flipped, was first observed in the hadronic $\Lambda$ production at Fermilab in 1976 as a surprise~\cite{Bunce:1976yb}. Since then large SSAs (or other related spin effects) have been consistently observed in various experiments at different collision energies~\cite{review}, such as Sivers and Collins asymmetries in the semi-inclusive hadron production in deep inelastic scattering (SIDIS) and in hadronic collisions~\cite{sidis, SSA-rhic}, as well as the large $\cos(2\phi)$ anomalous azimuthal asymmetry in back-to-back dihadron production in $e^+e^-$ annihilation~\cite{belle}.

To understand all these non-trivial and interesting spin effects and to explore the physics behind these asymmetries, two QCD-based approaches have been proposed and widely applied in the phenomenological studies~\cite{Anselmino:2008sga, DY, Anselmino:2007fs, newpion, Kanazawa:2010au, Kang:2008qh, Anselmino}: the collinear twist-three factorization approach~\cite{twist-three} and the transverse momentum dependent (TMD) factorization approach~\cite{Siv90, Collins93, mulders, boer-mulders, bmp}. These two approaches apply in different kinematic domain, and have been shown to be equivalent in the overlap region where they both apply; thus provide a unified QCD description for these spin effects~\cite{unify}. In the collinear twist-three factorization approach, the spin effect depends on certain twist-three multi-parton correlation functions. On the other hand, in the TMD factorization approach, the spin effect could be described in terms of TMD distributions and fragmentation functions. These twist-three multi-parton correlation functions are closely related to the TMD functions. For example, so-called Efremov-Teryaev-Qiu-Sterman quark-gluon correlation function is the first transverse-momentum-moment of the Sivers function~\cite{bmp}.

The spin effect could be generated from either the spin correlation in the parton distribution functions, among which the Sivers~\cite{Siv90} and Boer-Mulders~\cite{boer-mulders} functions are the important examples; or the spin correlation in the fragmentation functions, among which the Collins function~\cite{Collins93} and polarizing fragmentation function~\cite{mulders} are the important ones. Although all these four functions are naive-time-reversal-odd (T-odd), they have very different universality properties. For both Sivers and Boer-Mulders functions, it has been shown that they differ by a sign for the SIDIS and Drell-Yan (DY) processes~\cite{Col02}. On the other hand, both Collins function and the polarizing fragmentation function have been argued to be universal between different processes \cite{Metz:2002iz,Meissner:2008yf,Gamberg:2008yt,collins-s,Yuan:2009dw,Boer:2010ya}. The different universality properties are connected with the non-trivial initial- and final-state interactions between the active parton and the target remnant~\cite{Brodsky:2002cx}, whose interesting consequences remain to be tested in the future experiments~\cite{DY, Boer:2010ya}.  

On the fragmentation side, both Collins function and polarizing fragmentation function have been widely used in describing the spin asymmetries observed in the experiments. Collins function describes the transversely polarized quark jet fragmenting into an unpolarized hadron, whose transverse momentum relative to the jet axis correlates with the transverse polarization of the fragmenation quarks. It has been believed to be responsible for the azimuthal asymmetry $A_N^{\sin(\phi_h+\phi_s)}$ observed in SIDIS~\cite{sidis}, and the $\cos(2\phi)$ asymmetry observed in back-to-back dihadron production in $e^+e^-$ annihilations~\cite{belle}. On the other hand, polarizing fragmentation function describes the distribution of transversely polarized hadron in an unpolarized quark, through a correlation between their relative transverse momentum and the hadron transverse spin vector, which have been believed to be responsible for the Hyperon polarization observed in the experiments~\cite{hyperon}. 

The first transverse-momentum-moment of the Sivers and Boer-Mulders functions corresponds to twist-three quark-gluon correlation functions $T_{q, F}(x, x)$ and $T^{(\sigma)}_{q, F}(x, x)$~\cite{bmp}, for which the QCD evolution equations have been studied by various authors~\cite{Kang:2008ey, Zhou:2008mz, Vogelsang:2009pj, Braun:2009mi}. On the other hand, the corresponding fragmentation correlation functions connected to the first transverse-momentum-moment of the Collins function and the polarizing fragmentation function have been identified only recently~\cite{Yuan:2009dw, Boer:2010ya}. The purpose of our paper is to derive the QCD evolution equations for these relevant fragmentation correlation functions. QCD evolution equations are important in the sense that they control the energy dependence of the associated spin observables, also they enable us to evaluate the higher-order corrections to the spin-dependent cross sections systematically. 

\section{Evolution equations}
To start, we recall that the eight leading-twist TMD fragmentation functions could be defined from the following correlator
\ben
\Delta(z_h, p_\perp)&=&\frac{1}{z_h}\int\frac{d\xi^-d^2\xi_\perp}{(2\pi)^3}e^{ik^+\xi^{-}-i\vec{k}_\perp\cdot \vec{\xi}_\perp }
\langle 0| {\cal L}_\xi\psi(\xi)|P S X\rangle
\langle P S X| \bar{\psi}(0){\cal L}_0^\dagger|0\rangle,
\een
where a factor $1/2N_c$ from the average over the spin and the color of the fragmenting quark is suppressed. $P$ is the momentum of the final-state hadron with spin $S$, which has a transverse component $p_\perp$ relative to the momentum $k$ of the fragmenting quark. We choose the hadron to move along the $+z$ direction, and define two light-like vectors:
\ben
\bar{n}^\mu=\left[1^+, 0^-, 0_\perp\right], \qquad
n^\mu=\left[0^+, 1^-, 0_\perp\right].
\een
The momentum fraction $z_h=P^+/k^+$, and $\vec{k}_\perp=-\vec{p}_\perp/z_h$.
To ensure gauge invariance, we have explicitly written out the gauge link ${\cal L}_\xi={\cal P}\exp\left(-ig\int_0^\infty d\lambda n\cdot A(\xi+\lambda n)\right)$. 

The correlator $\Delta(z_h, p_\perp)$ could be expanded as follows~\cite{mulders, boer-mulders}
\ben
\Delta(z_h, p_\perp)&=&\frac{1}{2}
\Bigg[D_1(z_h, p_\perp^2)\sla{\bar{n}}
+\lambda_h G_{1L}(z_h, p_\perp^2)\gamma^5 \sla{\bar{n}}
+H_1(z_h, p_\perp^2) i\sigma_{\alpha\beta}\gamma^5 \bar{n}^\alpha S_{\perp}^\beta
\nnu
&&
+D_{1T}^\perp(z_h, p_\perp^2)\frac{\epsilon_{\alpha\beta\rho\sigma}\gamma^\alpha \bar{n}^\beta p_\perp^\rho S_{\perp}^\sigma}{M_h}
+H_1^\perp(z_h, p_\perp^2)\frac{\sigma_{\alpha\beta}p_\perp^\alpha\bar{n}^\beta}{M_h}
\nnu
&&
+G_{1T}(z_h, p_\perp^2)\frac{\vec{p}_\perp\cdot \vec{S}_{\perp}}{M_h}\gamma^5 \sla{\bar{n}}
+\lambda_h H_{1L}^\perp(z_h, p_\perp^2) \frac{i\sigma_{\alpha\beta}\gamma^5 \bar{n}^\alpha p_\perp^\beta}{M_h}
\nnu
&&
\left.
+H_{1T}^\perp(z_h, p_\perp^2)\frac{\vec{p}_\perp\cdot \vec{S}_{\perp}p_\perp^\beta -\frac{1}{2}\vec{p}_\perp^{\,2} S_{\perp}^\beta}{M_h^2} i \sigma_{\alpha\beta}\gamma^5 \bar{n}^\alpha
\right],
\een
where $\lambda_h$ is the hadron helicity and $M_h$ is the hadron mass. Out of the above eight TMD fragmentation functions, $D_1(z_h, p_\perp^2)$, $G_{1L}(z_h, p_\perp^2)$, and $H_1(z_h, p_\perp^2)$ are $p_\perp$-even functions and correspond to the unpolarized, longitudinally and transversely polarized distributions in the fragmentation. They are related to the leading-twist collinear fragmentation functions after integrating over $p_\perp$, for which the QCD evolution equations have been well-known~\cite{dglap, transversity}. All other five TMD fragmentation functions are $p_\perp$-odd functions in the sense that they vanish if integrate over $p_\perp$ directly. However, for $D_{1T}^\perp(z_h, p_\perp^2)$, $H_1^\perp(z_h, p_\perp^2)$, $G_{1T}(z_h, p_\perp^2)$, and $H_{1L}^\perp(z_h, p_\perp^2)$, their integral over $p_\perp$ after first weighted by $p_\perp^2$ (called ``first $p_\perp$-moment'') lead to the twist-three collinear fragmentation correlations; for $H_{1T}^\perp(z_h, p_\perp^2)$, the non-vanishing integral needs to be weighted by even higher power $p_\perp$ and actually corresponds to twist-four fragmentation correlation. 

In this paper, we are particularly interested in deriving the QCD evolution equations for the first transverse-momentum-moment of $H_1^\perp(z_h, p_\perp^2)$ and $D_{1T}^\perp(z_h, p_\perp^2)$:  $H_1^\perp(z_h, p_\perp^2)$ is the Collins function, and $D_{1T}^\perp(z_h, p_\perp^2)$ is the polarizing fragmentation function, which have been the main focus in the phenomenological studies. Their first transverse-momentum-moments have been identified recently, and given by the following fragmentation correlation functions~\cite{Yuan:2009dw, Boer:2010ya}
\ben
\hat{H}(z)&=&\frac{z^2}{2}\int\frac{d\xi^-}{2\pi}e^{ik^+\xi^{-}}
\frac{1}{2}\Big\{ {\rm Tr}\sigma^{\alpha\beta}n_\beta
\langle 0| \left[iD_{\perp\alpha}
+g\int_{\xi^-}^{\infty} d\eta^{-} F^{+}_{~~\alpha}(\eta^-)
\right]
\psi(\xi^-)|P S X\rangle
\langle P S X| \bar{\psi}(0)|0\rangle
+h.c.\Big\},
\\
\hat{T}(z)&=&z^2\int\frac{d\xi^-}{2\pi}e^{ik^+\xi^{-}}
\frac{1}{2}\Big\{ {\rm Tr}\,\sla{n}
\langle 0|\epsilon^{n\bar{n} S_{\perp} \alpha} \left[iD_{\perp\alpha}
+g\int_{\xi^-}^{\infty} d\eta^{-} F^{+}_{~~\alpha}(\eta^-)
\right]
\psi(\xi^-)|P S X\rangle
\langle P S X| \bar{\psi}(0)|0\rangle
+h.c.\Big\},
\een
where $D_\perp^\alpha=\partial_\perp^\alpha - ig A_\perp^\alpha$ is the covariant derivative, $F^{\alpha\beta}$ is the gluon field strength tensor. $\hat H(z)$ and $\hat T(z)$ are related to the Collins function $H_1^\perp(z_h, p_\perp^2)$ and polarizing fragmentation function $D_{1T}^\perp(z_h, p_\perp^2)$ as follows:
\ben
\hat H(z)&=&\int d^2p_\perp \frac{|\vec{p}_\perp|^2}{M_h} H_1^\perp(z, p_\perp^2),
\\
\hat T(z)&=&\int d^2p_\perp \frac{|\vec{p}_\perp|^2}{M_h} D_{1T}^\perp(z, p_\perp^2).
\een
According to Refs.~\cite{Yuan:2009dw, Boer:2010ya, Ji:1993vw}, the above defined {\it one-variable} fragmentation correlations belong to the more general twist-three {\it two-variable} fragmentation correlations, which are defined as
\ben
\hat H_D(z_1, z_2) &=& \frac{z_1 z_2}{2} \int\frac{d\xi^-}{2\pi}\frac{d\eta^-}{2\pi}
e^{ik_2^+\xi^{-}} e^{ik_g^+ \eta^-} 
\frac{1}{2}\Big\{ {\rm Tr}\,\sigma^{\alpha\beta}n_\beta
\langle 0|iD_{\perp\alpha}(\eta^-)
\psi(\xi^-)|P S X\rangle
\langle P S X| \bar{\psi}(0)|0\rangle
+h.c.\Big\},
\label{hd}
\\
\hat T_D(z_1, z_2) &=& z_1 z_2 \int\frac{d\xi^-}{2\pi}\frac{d\eta^-}{2\pi}
e^{ik_2^+\xi^{-}} e^{ik_g^+ \eta^-} 
\frac{1}{2}\Big\{ {\rm Tr}\,\sla{n}
\langle 0|\epsilon^{n\bar{n} S_{\perp} \alpha} iD_{\perp\alpha}(\eta^-)
\psi(\xi^-)|P S X\rangle
\langle P S X| \bar{\psi}(0)|0\rangle
+h.c.\Big\},
\label{td}
\een
where $k_g^+=k_1^+-k_2^+$ with $k_1^+=P^+/z_1$ and $k_2^+=P^+/z_2$. Similarly, one can define the corresponding $F$-type fragmentation correlations $\hat H_F(z_1, z_2)$ and $\hat T_F(z_1, z_2)$ by replacing $D_{\perp}^\alpha$ by $gF^{+\alpha}$ in Eqs.~(\ref{hd}) and (\ref{td}). By using the equation of motion, $D$-type and $F$-type functions are related to each other~\cite{bmp, Yuan:2009dw, Eguchi:2006qz}
\ben
\hat H_D(z_1, z_2)&=& {\rm PV}\left(\frac{1}{\frac{1}{z_1}-\frac{1}{z_2}}\right)\hat H_F(z_1, z_2)+\delta\left(\frac{1}{z_1}-\frac{1}{z_2}\right)\frac{z_1}{z_2}\hat H(z_2),
\\
\hat T_D(z_1, z_2)&=& {\rm PV}\left(\frac{1}{\frac{1}{z_1}-\frac{1}{z_2}}\right)\hat T_F(z_1, z_2)+\delta\left(\frac{1}{z_1}-\frac{1}{z_2}\right)\frac{z_1}{z_2}\hat T(z_2),
\een
where PV stands for principle value. The above equations show that $\hat H_D$ and $\hat T_D$ are more singular than their corresponding $F$-type functions $\hat H_F$ and $\hat T_F$ at $z_1=z_2$. Moreover, recent study has shown that $\hat H_F(z_1, z_2)$ and $\hat T_F(z_1, z_2)$ vanish when $z_1=z_2$~\cite{Meissner:2008yf, Gamberg:2008yt}. Therefore, it is more convenient to use the set of $\left\{\hat H(z), \hat H_F(z_1, z_2)\right\}$ and $\left\{\hat T(z), \hat T_F(z_1, z_2)\right\}$ when we derive the evolution equations, as we will follow below.
\bef
\psfig{file=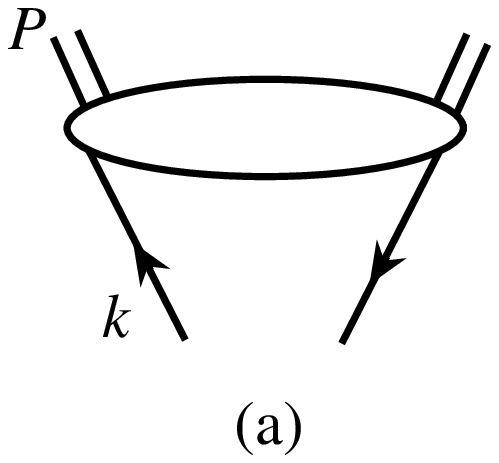, width=1.0in}
\hskip 0.3in
\psfig{file=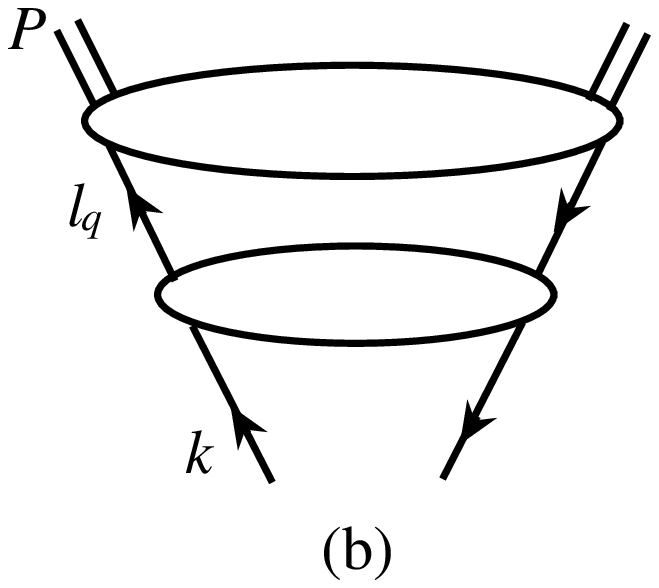, width=1.2in}
\hskip 0.3in
\psfig{file=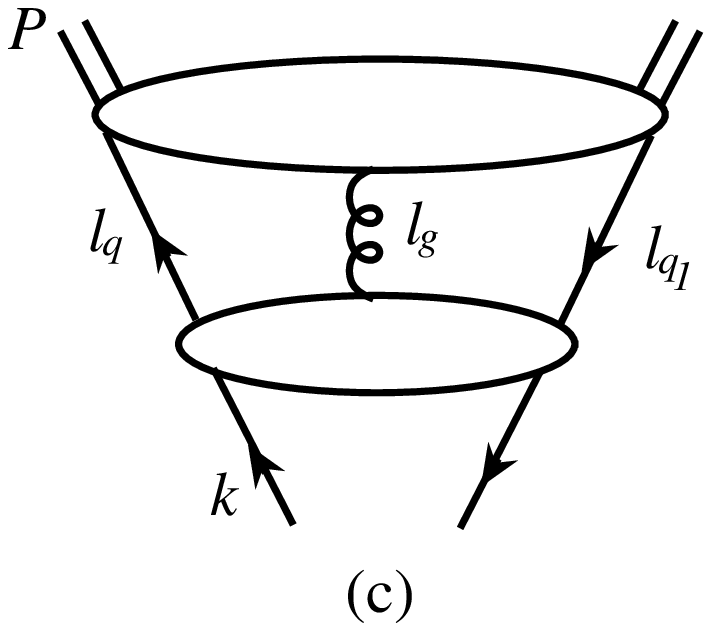, width=1.2in}
\caption{Feynman diagram representation: (a) the first transverse-momentum-moment of TMD fragmentation functions, (b) evolution contribution from itself: $\ell_q\approx P/z+\ell_{q_\perp}$, (c) evolution contribution from the {\it two-variable} $F$-type fragmentation correlations: $\ell_q=P/z$, $\ell_{q_1}=P/z_1$, and $\ell_g=\ell_q-\ell_{q_1}$.}
\label{general}
\eef

The fragmentation correlations $\hat H(z)$ and $\hat T(z)$ could be represented by the same cut forward scattering diagrams as sketched in Fig.~\ref{general}(a), but with different cut vertices which are used to connect the operator definition of the fragmentation correlations to the cut forward scattering Feynman diagrams. The standard way to derive the cut vertex is to express the operator definition of the correlation functions in terms of hadronic matrix elements of quark and gluon operators in momentum space, for details, see~\cite{Kang:2008ey}. The cut vertices for $\hat H(z)$ and $\hat T(z)$ as represented in Fig.~\ref{general}(a) are given by
\ben
\hat H(z): && \frac{z^2}{4}\delta\left(k^+ - \frac{P^+}{z}\right) i\,\gamma\cdot k_\perp\gamma\cdot n,
\label{cutH}
\\
\hat T(z): && \frac{z^2}{2}\delta\left(k^+ - \frac{P^+}{z}\right) \gamma\cdot n \epsilon^{n\bar{n} S_{\perp} k_\perp},
\label{cutT}
\een
where $k_\perp$ is the transverse momentum of the fragmenting quark with respect to the final hadron momentum.

Let's now explain how to derive the evolution equations. We will work in the light-cone gauge $n\cdot A=0$. In order to derive the evolution equations and evolution kernels from the operator definitions of the fragmentation correlations, one needs to compute the perturbative modification to these functions caused by the quark-gluon interactions in QCD. For both $\hat H(z)$ and $\hat T(z)$, the perturbative modification could come from either these correlation functions themselves, as shown in Fig.~\ref{general}(b); or the corresponding $F$-type fragmentation correlations, as shown in Fig.~\ref{general}(c). Since both $\hat H(z)$ and $\hat T(z)$ correspond to the operator $\sim \langle 0| \partial_\perp\psi|PSX\rangle\langle PSX|\bar{\psi}|0\rangle$, i.e., the partial derivative in the quark field, in order to calculate the contribution from themselves as shown in Fig.~\ref{general}(b), one has to perform collinear expansion. In other words, one should assume $\ell_q\approx P/z+\ell_{q_\perp}$, and the linear in $\ell_{q_\perp}$ expansion term when combined with the quark field from the  top blob will lead to the fragmentation correlations $\hat H(z)$ and $\hat T(z)$. On the other hand, to calculate the contribution from the $F$-type fragmentation correlations as shown in Fig.~\ref{general}(c), one has to insert a $A_\perp$ gluon in the Feynman diagram and then convert $A_\perp^\alpha$ to field strength $F^{+\alpha}$ through a partial integration, which leads to the $F$-type fragmentation correlations. In the calculation of $A_\perp$ contributions, since no collinear expansion is involved, one could set all the parton momenta as collinear to the hadron: $\ell_q=P/z$, $\ell_{q_1}=P/z_1$, and $\ell_g=\ell_q-\ell_{q_1}$. To sum up, the perturbative modifications could be written as
\ben
d\hat H(z_h, \mu^2)&=&\int dz\frac{1}{2z^4}\frac{\partial}{\partial \ell_{q_\perp}^\alpha}{\rm Tr} \left[i\, \gamma^\alpha \gamma\cdot P {\mathcal K}(k, \ell_q\approx P/z+\ell_{q_\perp})\right]_{\ell_{q_\perp}\to 0}\hat H(z,\mu^2)
+\int dz dz_1 {\rm PV}\left(\frac{1}{\frac{1}{z}-\frac{1}{z_1}}\right)
\nnu
&&
\times
\frac{1}{2z^3 z_1^3}{\rm Tr}\left[i\, \gamma^\alpha \gamma\cdot P {\mathcal K}_\alpha(k, \ell_q=P/z, \ell_{q_1}=P/z_1)\right]\hat H_F(z, z_1,\mu^2),
\label{hz}
\\
d\hat T(z_h, \mu^2)&=& \int \frac{dz}{2z^4}\frac{\partial}{\partial \ell_{q_\perp}^\alpha}{\rm Tr} \left[\gamma\cdot P \epsilon^{\bar{n} n S_{\perp}\alpha} {\mathcal K}(k, \ell_q\approx P/z+\ell_{q_\perp}) \right]_{\ell_{q_\perp}\to 0}\hat T(z, \mu^2)
+\int dz dz_1 {\rm PV}\left(\frac{1}{\frac{1}{z}-\frac{1}{z_1}}\right)
\nnu
&&
\times
\frac{1}{2z^3 z_1^3}{\rm Tr}\left[\gamma\cdot P \epsilon^{\bar{n}n S_{\perp}\alpha}{\mathcal K}_\alpha(k, \ell_q=P/z, \ell_{q_1}=P/z_1)\right]\hat T_F(z, z_1, \mu^2),
\label{tz}
\een
where $\mu$ is the factorization scale, ${\mathcal K}(k, \ell_q\approx P/z+\ell_{q_\perp})$ and ${\mathcal K}_\alpha(k, \ell_q=P/z, \ell_{q_1}=P/z_1)$ are the hard partonic part calculated from Fig.~\ref{general}(b) and (c) without the top blob, respectively. Note we use the same symbols ${\mathcal K}(k, \ell_q\approx P/z+\ell_{q_\perp})$ and ${\mathcal K}_\alpha(k, \ell_q=P/z, \ell_{q_1}=P/z_1)$ for both $\hat H(z_h, \mu^2)$ and $\hat T(z_h, \mu^2)$ for simplicity, they are different in the calculations as shown below.
\bef
\psfig{file=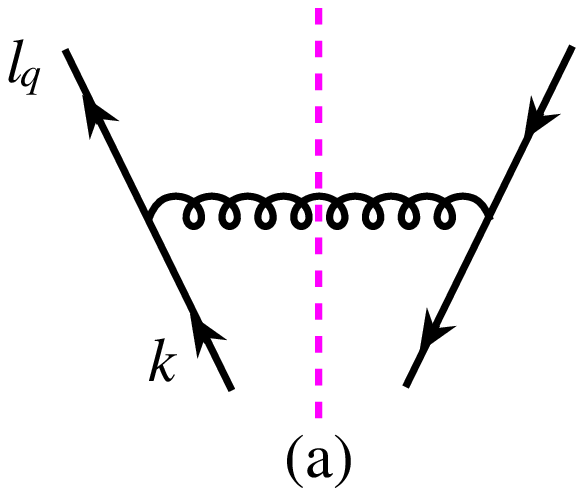, width=1.1in}
\hskip 0.2in
\psfig{file=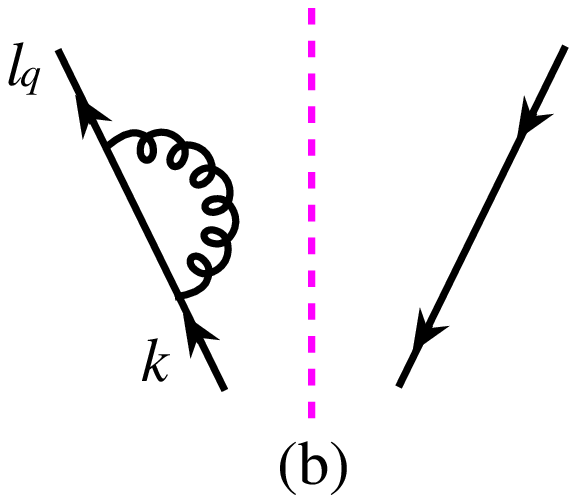, width=1.1in}
\hskip 0.2in
\psfig{file=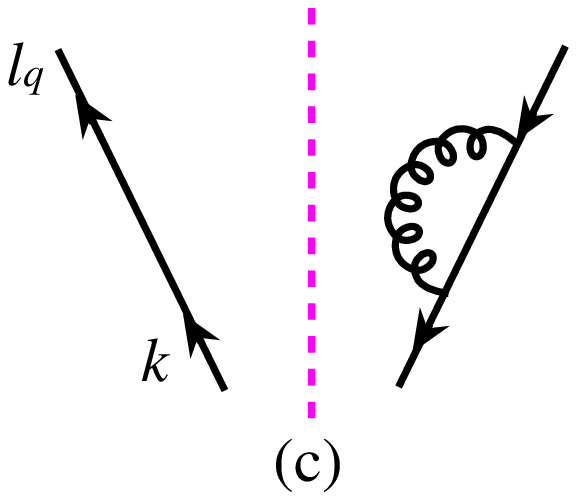, width=1.1in}
\caption{Contribution from the first transverse-momentum-moment of the TMD fragmentation functions themselves: (a) real contribution, (b) and (c) virtual contributions.}
\label{itself}
\eef

To the leading order in strong coupling constant $\alpha_s$, the contribution from the fragmentation correlation functions themselves  as in Fig.~\ref{general}(b) are given by the Feynman diagrams in Fig.~\ref{itself}: \ref{itself}(a) is the real contribution, while \ref{itself}(b) and (c) are the virtual contributions. These diagrams are the same as those when one calculates the evolution kernel for the leading-twist unpolarized collinear fragmentation function \cite{dglap, Collins:1988wj}, but the actual calculations are very different. As we have explained above, in our current calculations, a collinear expansion is needed to pick up the linear in $\ell_{q_\perp}$ terms which lead to $\hat H(z)$ and $\hat T(z)$; on the other hand, there is no collinear expansion involved in the calculations for the leading-twist collinear functions. Following the collinear expansion as specified in the first terms of Eqs.~(\ref{hz}) and (\ref{tz}), we obtain the contributions from the real diagram Fig.~\ref{itself}(a):
\ben
d\hat H(z_h, \mu^2)|_{\rm Fig.~\ref{itself}(a)}&=&\int^{\mu^2} \frac{dk_\perp^2}{k_\perp^2} \frac{\alpha_s}{2\pi}C_F \int \frac{dz}{z}
\frac{2\hat z}{1-\hat z}\hat H(z, \mu^2),
\\
d\hat T(z_h, \mu^2)|_{\rm Fig.~\ref{itself}(a)}&=&\int^{\mu^2} \frac{dk_\perp^2}{k_\perp^2} \frac{\alpha_s}{2\pi}C_F \int \frac{dz}{z}
\frac{1+\hat z^2}{1-\hat z}\hat T(z, \mu^2),
\een
where $\hat z=z_h/z$. For virtual diagrams Fig.~\ref{itself}(b) and (c), since now $\ell_q=k$, thus the $\ell_{q_\perp}$ expansion is in fact an expansion of $k_\perp$. On the other hand, the cut vertices for both $\hat H(z)$ and $\hat T(z)$ in Eqs.~(\ref{cutH}) and (\ref{cutT}) depend linearly on $k_\perp$, thus after a direct expansion over $k_\perp$ in the cut vertices, one could set all $k_\perp=\ell_{q_\perp}=0$ afterwards. The final results are
\ben
d\hat H(z_h, \mu^2)|_{\rm Fig.~\ref{itself}(b)+(c)}&=&-\int^{\mu^2} \frac{dk_\perp^2}{k_\perp^2} \frac{\alpha_s}{2\pi}C_F\int_0^1 \frac{dz'}{z'}
\frac{1+z'^2}{1-z'}\hat H(z_h, \mu^2),
\\
d\hat T(z_h, \mu^2)|_{\rm Fig.~\ref{itself}(b)+(c)}&=&-\int^{\mu^2} \frac{dk_\perp^2}{k_\perp^2} \frac{\alpha_s}{2\pi}C_F\int_0^1 \frac{dz'}{z'}
\frac{1+z'^2}{1-z'}\hat T(z_h, \mu^2).
\een
Combining above real and virtual contributions, we obtain
\ben
d\hat H(z_h, \mu^2)|_{\rm Fig.~\ref{itself}}&=&\int^{\mu^2} \frac{dk_\perp^2}{k_\perp^2} \frac{\alpha_s}{2\pi}C_F \int \frac{dz}{z}
\left[\frac{2\hat z}{(1-\hat z)_{+}} + \frac{3}{2}\delta(1-\hat z)\right]\hat H(z, \mu^2),
\\
d\hat T(z_h, \mu^2)|_{\rm Fig.~\ref{itself}}&=&\int^{\mu^2} \frac{dk_\perp^2}{k_\perp^2} \frac{\alpha_s}{2\pi}C_F \int \frac{dz}{z}
\left[\frac{1+\hat z^2}{(1-\hat z)_+}+\frac{3}{2}\delta(1-\hat z)\right]\hat T(z, \mu^2).
\een

Let's now consider the contribution from the $F$-type fragmentation correlation functions. To the leading order in $\alpha_s$, the relevant Feynman diagrams are shown in Figs.~\ref{real} and \ref{virtual}: the real contributions in Fig.~\ref{real} and the virtual contributions in Fig.~\ref{virtual}.
\bef
\psfig{file=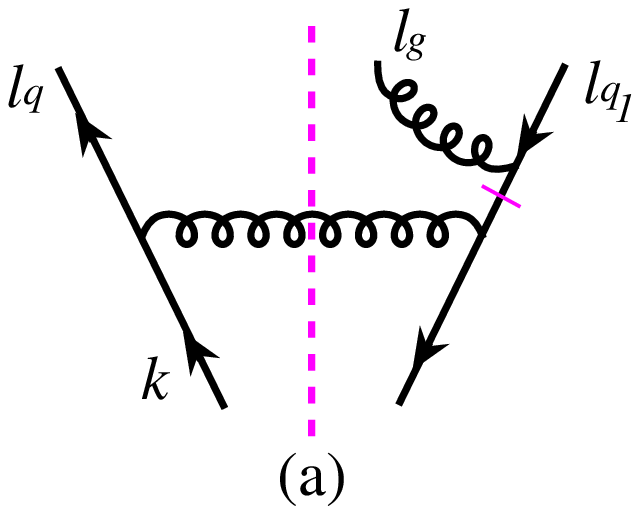, width=1.2in}
\hskip 0.2in
\psfig{file=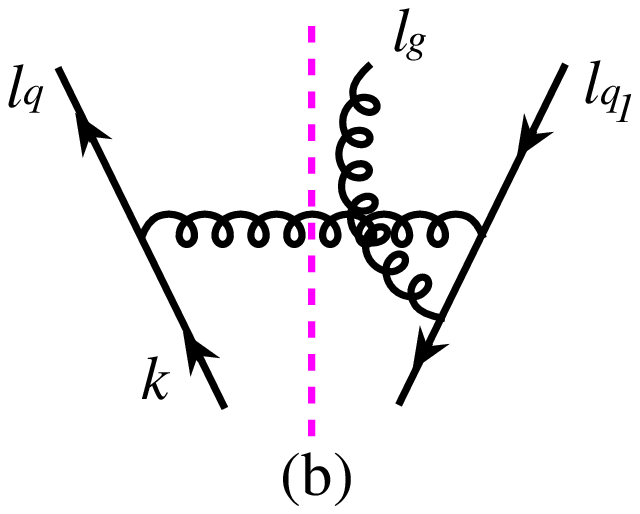, width=1.2in}
\hskip 0.2in
\psfig{file=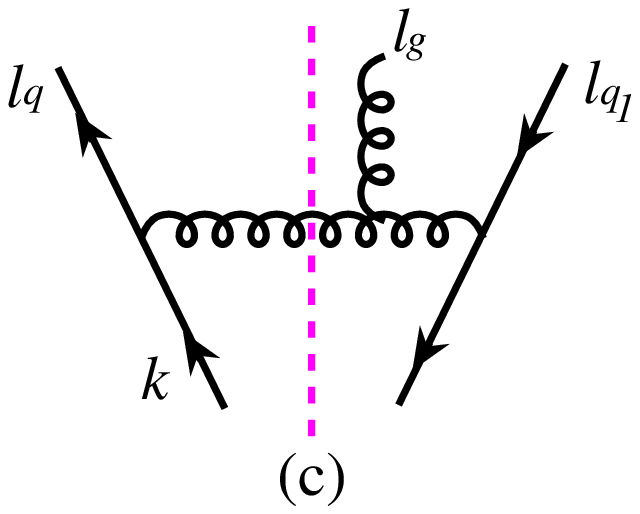, width=1.2in}
\caption{Contribution from the $F$-type fragmentation correlation functions: real diagrams. The ``mirror'' diagrams for which the additional gluon attaches on the left of the cut are not shown, but are included in the calculations.}
\label{real}
\eef
As we have mentioned above, calculating the $A_\perp$ contribution (finally related to $F$-type fragmentation correlation functions) does not involve collinear expansion in the light-cone gauge, we could set all the parton momenta collinear to the final hadron:
\ben
\ell_q=P/z, \qquad \ell_{q_1}=P/z_1, \qquad \ell_g=\ell_q-\ell_{q_1}.
\label{lqq1}
\een
The calculations following the formalism in the second terms of Eqs.~(\ref{hz}) and (\ref{tz}) are also straightforward. For the real diagram contributions, we collect the terms through the color factors: Fig.~\ref{real}(a), (b), and (c) have color factors $C_F$, $C_F-C_A/2$, and $C_A/2$, respectively. The final results are
\ben
d\hat{H}(z_h, \mu^2)|_{\rm Fig.~\ref{real}}
&=&\int^{\mu^2} \frac{dk_\perp^2}{k_\perp^2} \frac{\alpha_s}{2\pi}\int \frac{dz}{z}
\int \frac{dz_1}{z_1^2}{\rm PV}\left(\frac{1}{\frac{1}{z}-\frac{1}{z_1}}\right)
B(z_h, z, z_1)\hat H_F(z, z_1, \mu^2),
\\
d\hat{T}(z_h, \mu^2)|_{\rm Fig.~\ref{real}}
&=&\int^{\mu^2} \frac{dk_\perp^2}{k_\perp^2} \frac{\alpha_s}{2\pi}\int \frac{dz}{z}
\int \frac{dz_1}{z_1^2}{\rm PV}\left(\frac{1}{\frac{1}{z}-\frac{1}{z_1}}\right)
B'(z_h, z, z_1)\hat T_F(z, z_1, \mu^2),
\een
where the kernel $B(z_h, z, z_1)$ and $B'(z_h, z, z_1)$ are given by
\ben
B(z_h, z, z_1)& = & C_F \left[\frac{2z_h}{z}\left(1+\frac{z_h}{z_1}-\frac{z_h}{z}\right)
\right]
+
\frac{C_A}{2}\left[\frac{2z_h}{z}
\frac{z_h^2(z^2+z_1^2)-z z_1 (z+z_1)}{(z_1-z)(z_1-z_h)z}
\right],
\label{bz}
\\
B'(z_h, z, z_1)& = &C_F \left[\frac{z_h}{z_1}-\frac{z}{z_1}-\frac{z_h}{z}+\frac{z_h^2}{z z_1}+2
\right]
+
\frac{C_A}{2}\left[\frac{(z z_h+z_1z_h -2 z z_1)(z z_1+z_h^2)}{(z_1-z)(z_1-z_h)z^2}
\right].
\label{bpz}
\een

Finally let's consider the virtual contributions from the $F$-type fragmentation correlation functions as shown in Fig.~\ref{virtual}. It is important to realize that for all the diagrams (a)-(e) in Fig.~\ref{virtual}, we have (follow Eq.~(\ref{lqq1}))
\ben
k=\ell_q=P/z,
\een
which has no transverse component, i.e., $k_\perp=0$. Note the cut vertices used to define both $\hat H(z)$ and $\hat T(z)$ depend linearly on $k_\perp$, see Eqs.~(\ref{cutH}) and (\ref{cutT}). Thus when $k_\perp=0$, they vanish. In other words, all these virtual diagrams do not contribute. Thus the perturbative modifications for $\hat H(z_h, \mu^2)$ and $\hat T(z_h, \mu^2)$ receive contributions from only Figs.~\ref{itself} and \ref{real}. Adding them up, we obtain
\ben
d\hat{H}(z_h, \mu^2)
&=&\int^{\mu^2} \frac{dk_\perp^2}{k_\perp^2}
\frac{\alpha_s}{2\pi}\int \frac{dz}{z}
\left[A(\hat{z}) \hat{H}(z, \mu^2)
+\int \frac{dz_1}{z_1^2}{\rm PV}\left(\frac{1}{\frac{1}{z}-\frac{1}{z_1}}\right)B(z_h, z, z_1) \hat H_F(z, z_1, \mu^2)
\right],
\\
d\hat T(z_h, \mu^2)
&=&\int^{\mu^2} \frac{dk_\perp^2}{k_\perp^2}
\frac{\alpha_s}{2\pi}\int \frac{dz}{z}
\left[A'(\hat{z}) \hat T(z, \mu^2)
+\int \frac{dz_1}{z_1^2}{\rm PV}\left(\frac{1}{\frac{1}{z}-\frac{1}{z_1}}\right)B'(z_h, z, z_1) \hat T_F(z, z_1, \mu^2)
\right].
\een
\bef
\psfig{file=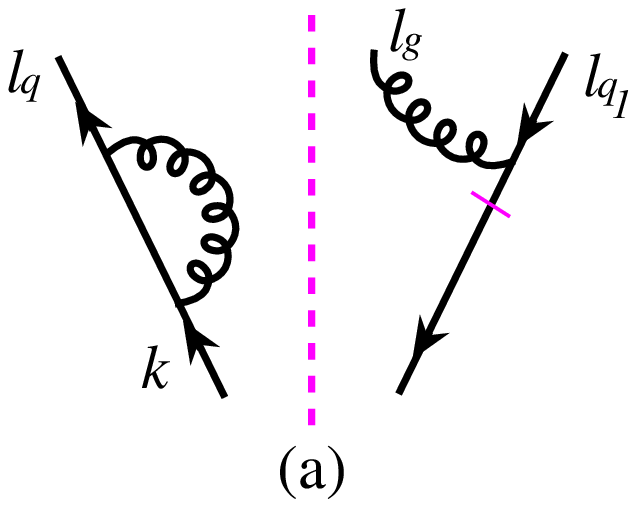, width=1.18in}
\hskip 0.05in
\psfig{file=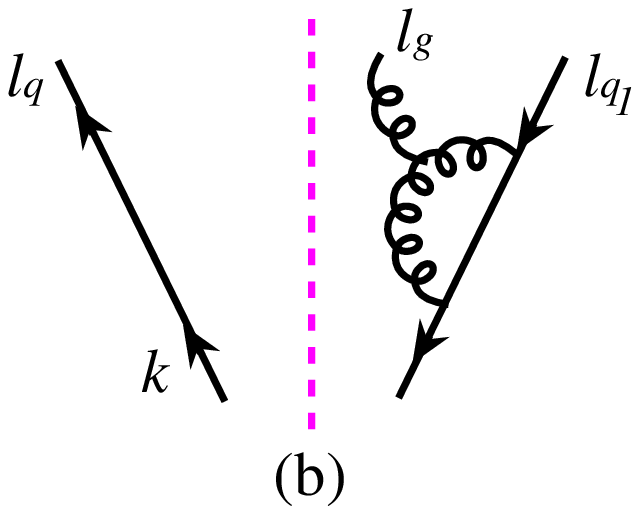, width=1.18in}
\hskip 0.05in
\psfig{file=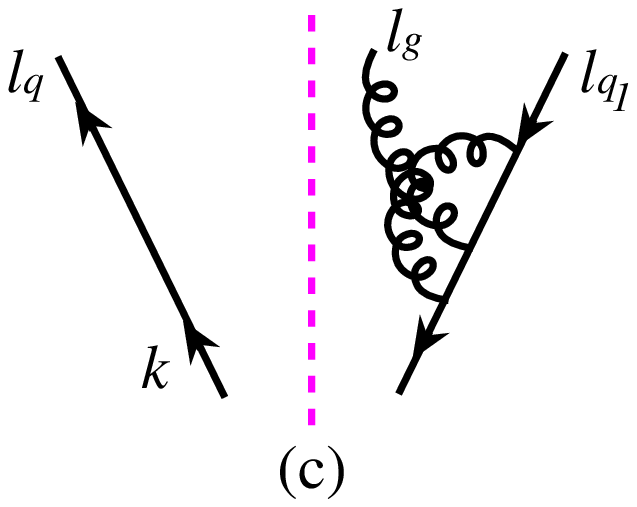, width=1.18in}
\hskip 0.05in
\psfig{file=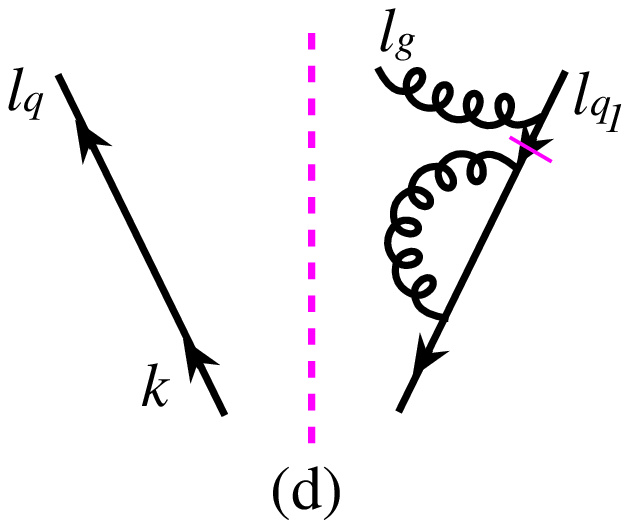, width=1.18in}
\hskip 0.05in
\psfig{file=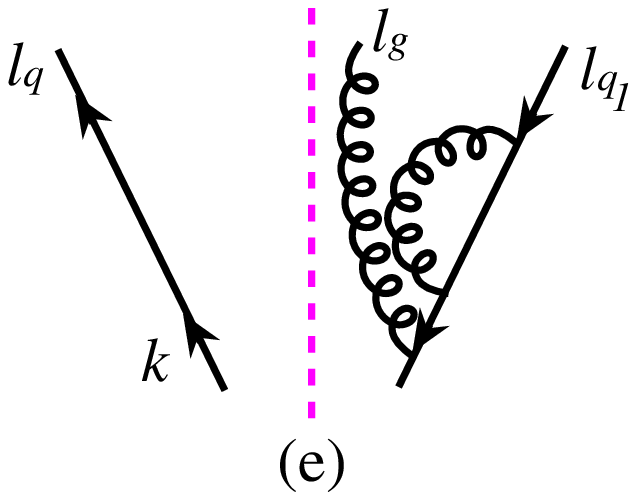, width=1.18in}
\caption{Contribution from the $F$-type fragmentation correlation functions: virtual diagrams. The ``mirror'' diagrams for which the additional gluon attaches on the left of the cut are not shown, but are included in the calculations.}
\label{virtual}
\eef
Differentiate both sides of above equations with respect to $\ln \mu^2$, we obtain the scale evolution equations for $\hat{H}(z_h, \mu^2)$ and $\hat{T}(z_h, \mu^2)$ as 
\ben
\frac{\partial \hat{H}(z_h, \mu^2)}{\partial \ln\mu^2} 
&=&\frac{\alpha_s}{2\pi}\int \frac{dz}{z}
\left[A(\hat{z}) \hat{H}(z, \mu^2)
+\int \frac{dz_1}{z_1^2}{\rm PV}\left(\frac{1}{\frac{1}{z}-\frac{1}{z_1}}\right)B(z_h, z, z_1) \hat H_F(z, z_1, \mu^2)
\right],
\label{evlH}
\\
\frac{\partial \hat T(z_h, \mu^2)}{\partial \ln\mu^2} 
&=&\frac{\alpha_s}{2\pi}\int \frac{dz}{z}
\left[A'(\hat{z}) \hat T(z, \mu^2)
+\int \frac{dz_1}{z_1^2}{\rm PV}\left(\frac{1}{\frac{1}{z}-\frac{1}{z_1}}\right)B'(z_h, z, z_1) \hat T_F(z, z_1, \mu^2)
\right],
\label{evlT}
\een
where $B(z_h, z, z_1)$ and $B'(z_h, z, z_1)$ are given in Eqs.~(\ref{bz}) and (\ref{bpz}), and $A(\hat{z})$ and $A'(\hat{z})$ have the following forms
\ben
A(\hat{z})&=&C_F\left[\frac{2\hat{z}}{(1-\hat{z})_+}+\frac{3}{2}\delta(1-\hat{z})\right],
\\
A'(\hat{z})&=&C_F\left[\frac{1+\hat{z}^2}{(1-\hat{z})_+}+\frac{3}{2}\delta(1-\hat{z})\right].
\een
Eqs.~(\ref{evlH}) and (\ref{evlT}) are the main results of our paper. A few comments about these results are provided:
\begin{itemize}
\item The evolution equations derived here for both $\hat H(z_h, \mu^2)$ and $\hat T(z_h, \mu^2)$ are not a close set of equations, as stand in Eqs.~(\ref{evlH}) and (\ref{evlT}): the evolutions depend on the {\it diagonal} pieces $\hat H(z, \mu^2)$ and $\hat T(z, \mu^2)$, as well as the {\it off-diagonal} pieces $\hat H_F(z, z_1, \mu^2)$ and $\hat T_F(z, z_1, \mu^2)$. 
\item It is interesting to notice that the evolution kernel $A(\hat z)$ is the same as that for the transversity fragmentation function~\cite{transversity}, while the kernel $A'(\hat z)$ is the same as that for the unpolarized fragmentation function~\cite{dglap}. 
\item As shown in Refs.~\cite{Meissner:2008yf, Gamberg:2008yt}, both $\hat H_F(z, z_1, \mu^2)$ and $\hat T_F(z, z_1, \mu^2)$ vanish at $z=z_1$. This {\it might} imply that the {\it off-diagonal} pieces could be small, thus the evolution of $\hat H(z_h, \mu^2)$ might be close to that of transversity, while the evolution of $\hat T(z_h, \mu^2)$ might be close to that of unpolarized fragmentation function. If this were true, it will have important consequences on the current global analysis of the spin asymmetries \cite{Anselmino:2008sga, Anselmino:2007fs}. Of course, whether the {\it off-diagonal} terms play a less important role in determining the evolution of the {\it diagonal} terms needs to be tested from experimental data through global analysis, such as those done in \cite{newpion, Kanazawa:2010au}. 
\item It is also important to realize that there is no gluon Collins function, thus $\hat H(z_h, \mu^2)$ does not receive contribution from gluon part. On the other hand, there could be gluon polarizing fragmentation functions \cite{Bomhof:2006ra}, from which the corresponding gluon fragmentation correlation functions could be defined. Thus there could be contributions from the gluon part to the evolution of $\hat T(z_h, \mu^2)$, and these contributions are not studied here. 
\end{itemize}

\section{Conclusions}
We have derived the QCD evolution equations for the first transverse-momentum-moment of the naive-time-reversal-odd transverse momentum dependent fragmentation 
functions: the Collins function $H_1^\perp(z_h, p_\perp^2)$ and the polarizing fragmentation function $D_{1T}^\perp(z_h, p_\perp^2)$. These first transverse-momentum-moments correspond to twist-three fragmentation correlation functions denoted as $\hat H(z_h, \mu^2)$ and $\hat T(z_h, \mu^2)$. We calculate in light-cone gauge the order of $\alpha_s$ evolution kernel for the scale dependence of both $\hat H(z_h, \mu^2)$ and $\hat T(z_h, \mu^2)$. We find that the evolution of both fragmentation correlation functions receives contributions from themselves, as well as from the $F$-type {\it two-variable} fragmentation correlation functions $\hat H_F(z, z_1, \mu^2)$ and $\hat T_F(z, z_1, \mu^2)$. We find for $\hat H(z_h, \mu^2)$ that the {\it diagonal} piece in the evolution kernel is the same as that for the transversity fragmentation function, while for $\hat T(z_h, \mu^2)$ that the {\it diagonal} piece is the same as that for the unpolarized fragmentation function. Since the {\it off-diagonal} pieces involve the $F$-type fragmentation correlation functions which vanish at $z=z_1$, thus they might play a less important role. If this were true, it will provide important consequences in the current global analysis of spin asymmetries.

\section*{Acknowledgments}
We are grateful to RIKEN, Brookhaven National Laboratory, 
and the U.S. Department of Energy 
(Contract No.~DE-AC02-98CH10886) for supporting this work. 
We thank Institute for Nuclear Theory at University of Washington for its
hospitality during the writing of this work.


\end{document}